\documentclass{article}
\usepackage[top=1in, bottom=1in, left=1in, right=1in]{geometry}
\usepackage{amsmath,amsfonts,amscd,amssymb}
\usepackage{mathtools}
\usepackage{siunitx}
\usepackage{subeqnarray}
\usepackage[normalem]{ulem}
\usepackage{biblatex}
\newcommand{\Av}{\mathbf{A}}
\newcommand{\Bv}{\mathbf{B}}
\newcommand{\Cv}{\mathbf{C}}
\newcommand{\Dv}{\mathbf{D}}
\newcommand{\Hv}{\mathbf{H}}
\newcommand{\Qv}{\mathbf{Q}}
\newcommand{\Rv}{\mathbf{R}}
\newcommand{\Uv}{\mathbf{U}}
\newcommand{\Vv}{\mathbf{V}}

\newcommand{\uv}{\mathbf{u}}
\newcommand{\xv}{\mathbf{x}}
\newcommand{\Yv}{\mathbf{Y}}
\newcommand{\Sigmav}{\boldsymbol{\Sigma}}

\usepackage[bottom,flushmargin,hang,multiple]{footmisc}
\usepackage{lipsum}

\title{\vspace{-.55in}{\LARGE\textbf{Data-driven unsteady aeroelastic modeling for control}}\vspace{-.15in}}

\author{\normalsize{Michelle K. Hickner$^{1*}$, Urban Fasel$^1$, Aditya G. Nair, Bingni W. Brunton$^3$, and Steven L. Brunton$^1$}\\
\footnotesize{$^1$ Department of Mechanical Engineering, University of Washington, Seattle, WA 98195, United States}\\
\footnotesize{$^2$ Department of Mechanical Engineering, University of Nevada, Reno, NV 89557, United States}\\
\footnotesize{$^3$ Department of Biology, University of Washington, Seattle, WA 98195, United States\vspace{-.2in}}
}
\date{}

\begin{document}

\maketitle

\vspace{-.2in}
\begin{abstract} 
    Aeroelastic structures, from insect wings to wind turbine blades, experience transient unsteady aerodynamic loads that are coupled to their motion. 
    Effective real-time control of flexible structures relies on accurate and efficient predictions of both the unsteady aeroelastic forces and airfoil deformation. 
    For rigid wings, classical unsteady aerodynamic models have recently been reformulated in state-space for control and extended to include viscous effects. 
    Here we further extend this modeling framework to include the deformation of a flexible wing in addition to the quasi-steady, added mass, and unsteady viscous forces.  
    We develop low-order linear models based on data from direct numerical simulations of flow past a flexible wing at low Reynolds number. 
    We demonstrate the effectiveness of these models to track aggressive maneuvers with model predictive control while constraining maximum wing deformation. 
    This system identification approach provides an interpretable, accurate, and low-dimensional representation of an aeroelastic system that can aid in system and controller design for applications where transients play an important role.
    \\

\end{abstract}

\section*{Nomenclature}\label{symbols}
\vspace{-.05in}
\begin{tabular}{rl}
 $\alpha$ & angle of attack of plate \\
 $\kappa$ & curvature near leading edge of plate \\
 $\nu$ & kinematic viscosity \\
 $\tau$ & convective time (dimensionless), $\tau \equiv t U_{\infty}/c$ \\
 $\Delta \tau$ & time step\\
 $\Delta \tau_c$ & coarse time step, duration of impulse\\
 $a$ & pitch axis location relative to the $1/2$ chord\\
 $c$ & chord length of plate\\
 $\Delta c$ & length of discretized plate element \\
 $r$ & model rank\\
 $t$ & time \\
 $\xv$ & state vector of state-space model\\
 $\Av$, $\Bv$, $\Cv$ & state-space matrices identified by ERA\\
 $C_{\alpha}$ & angle of attack (lift slope) coefficient\\
 $C_{\dot{\alpha}}$, $C_{\ddot{\alpha}}$ & added mass coefficients\\
 $C_L$ & coefficient of lift\\
 $EI$ & plate flexural rigidity\\
 $\Hv$ & Hankel matrix, time-delay matrix of measurements\\
 $K_B$ & plate bending stiffness\\
 $M_{\rho}$ & mass ratio\\
 $\Qv$ & MPC weights for outputs \\
 $\Rv$ & MPC weights for inputs \\
 $Re$ & Reynolds number, $Re \equiv U_\infty c/\nu$ \\
 $\textbf{R}_{\Delta u}$ & MPC weights for input rate of change \\
 $T_c$ & MPC control horizon \\
 $T_p$ & MPC prediction horizon \\
 $U_{\infty}$ & free stream velocity\\
 $\Yv$ & matrix of measurements\\
 $\tilde{\Yv}$ & Markov parameters\\
\end{tabular}

\section{Introduction}
Wing flexibility plays a major role in aerodynamic systems, both natural and engineered, by increasing efficiency and mitigating structural damage from sudden loading~\cite{Fernandez-Gutierrez2021,Hang2021,Mountcastle2014,Reid2019}. 
The effect of wing flexibility has been long studied in the context of bio-locomotion~\cite{Daniel1988cjz,Fish1991mr,Dickinson1996jeb,Fish1996az,Dickinson:1999,Birch2001nature,Combes:2001,Sane2001jeb,Liao2003science,Tytell:2004,Lauder2005fp,Wang2005arfm,Hedenstrom2007science,Peng2008jeb,riskin2008quantifying,Dabiri2009arfm,Shelley2011arfm,Wu2011arfm,Leftwich2012jeb,Nawroth2012naturebio,Floryan2020}, where it is known that flexibility has favorable aerodynamic and hydrodynamic properties~\cite{Daniel2002,Mountcastle2013,Song2008,Tytell2016}. 

There is an increasing need to develop unsteady aeroelastic models to realize these benefits for engineering systems, for example to control autonomous flight vehicles, including at sizes comparable to insects and birds. 
The benefits of wing flexibility are often the consequences of subtle aeroelastic effects, and for many applications it is important to capture unsteady viscous fluid transients \cite{Tregidgo2013,Poirel2008}.  
This work develops accurate and efficient models to capture these effects in design and control efforts. 

Aeroelastic modeling is challenging because of the strongly coupled fluid-structure interactions, which are difficult to capture in reduced order models. 
Existing high-fidelity models can simulate the coupled fluid flow fields and the deformation profile of a structure~\cite{peskin2002,Mittal2005,Goza2017,Liu2014}, but they are computationally intensive and do not fit readily into model-based control design. 
Medium fidelity models, such as doublet-lattice methods~\cite{Albano1969}, strip theory~\cite{Kim2008,Taha2014}, and panel methods~\cite{Wang2021} are widely-used approaches that can be combined with optimization to aid in aeroelastic design~\cite{Fasel2021AIAA}, but they neglect viscous effects and are not well suited for real-time control. 
The unsteady vortex-lattice method provides some balance between accuracy and tractability for control; however, it has limitations for wings with significant camber or viscous effects~\cite{Hesse2014,Murua2012}. 
Other nonlinear modeling approaches include Volterra kernels~\cite{Brockett1976,Lucia2004, Balajewicz2012} and modal models~\cite{Lucia2005, Fonzi2020, Artola2021,Li2021}. Linear models are often sufficient for even large amplitude motions~\cite{Brunton2013_JFM} until leading edge separation and stall~\cite{Eldredge2019arfm}.
Considerable attention has been paid to development of aeroelastic models at or near transonic speeds~\cite{Hall2000,Thomas2003,Lucia2003,Lieu2006,Yang2020,Opgenoord2018,Geuzaine2003aeroelastic}, while techniques for small flexible wings at low Reynolds number have been gaining more interest recently~\cite{Tiomkin2021,Stanford2007,Kurdila1999,Liu2021,Bryant2013}. Use of these low Reynolds number modeling techniques is necessary for characterizations of micro air vehicles (MAVs) or investigation of biological flight~\cite{Shyy2013}. 

There are many existing techniques for developing reduced order models of aeroelastic systems that take advantage of data-driven system identification methods~\cite{Kou2021}, such as those based on reduced order bases from eigen-decompositions~\cite{Dowell2001,Dowell1997}, proper orthogonal decomposition (POD)~\cite{Amsallem2010,Lucia2004,Barone2005,Lieu2006}, dynamic mode decomposition (DMD), or the eigensystem realization algorithm (ERA)~\cite{Juang1985,Silva2001,Kim2008,Silva2008,Liu2021}. 
Reduced order state-space models have been developed for the control of lift, including those with significant viscous and unsteady effects, extending the classic Theodorsen~\cite{Theodorsen1935} and Wagner models~\cite{Wagner1925} based on viscous flow data~\cite{ Brunton2013_JFM,Brunton2013_JFS,Brunton2014,Hemati2017,Nardini2018}. 
These linear unsteady aerodynamic models use techniques such as ERA and quasi-linear parameter varying  techniques (qLPV) to develop minimal realizations that are both accurate and computationally efficient enough for real-time control. 
Data-driven techniques avoid the simplifying assumptions built into some analytical models, such as inviscid or quasi-steady flow, allowing models to be used in a wider range of flow conditions, including low Reynolds number flows and those resulting from rapid, large-amplitude maneuvers. 

\begin{figure*}
\vspace{-.1in}
\centering
\includegraphics[width=1\textwidth]{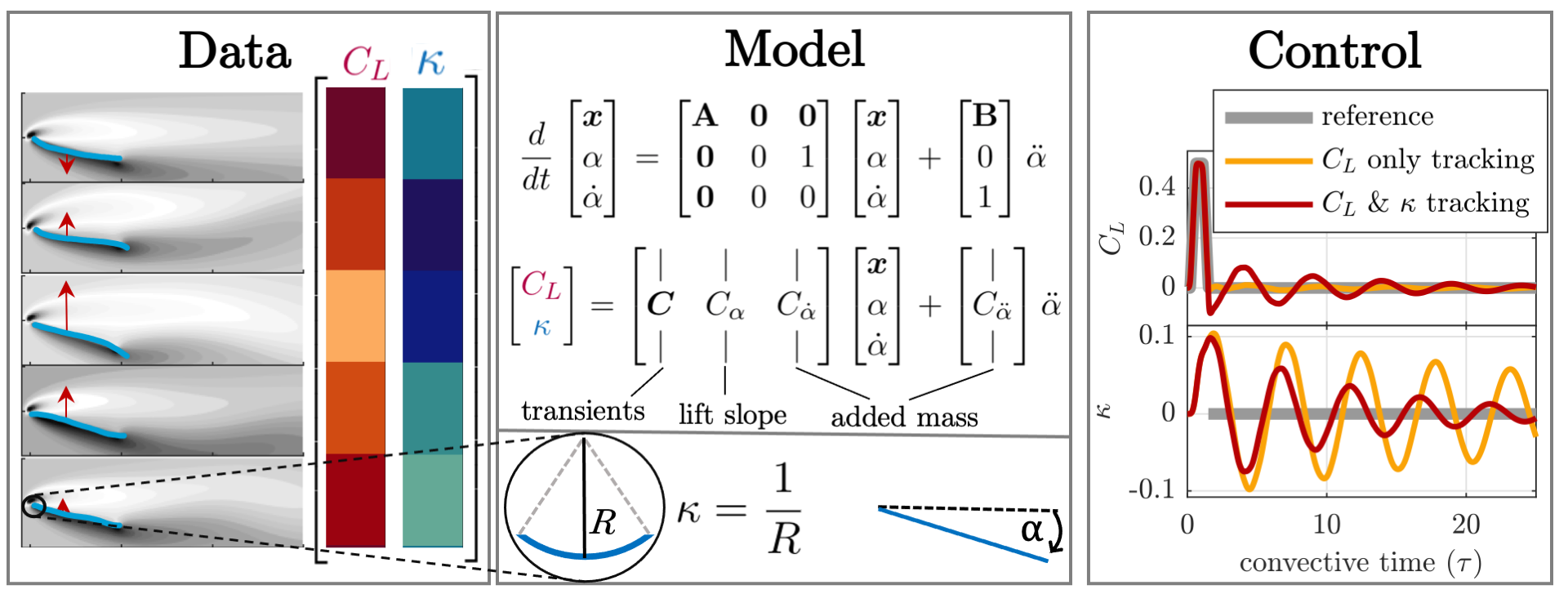}
\vspace{-.1in}
\caption{A linear state-space model of an aeroelastic wing is developed from lift and deformation observations. The model separates the contributions from transients, quasi-steady pitch effects, and added mass. The model is used to track reference $C_L$ and curvature, $\kappa$.}
\label{fig:overview}
\end{figure*}

In this work, we extend these rigid low-order unsteady aerodynamic models to include the effect of wing flexibility and demonstrate these models for the control of aggressive maneuvers.  
The resulting linear state-space models are purely data-driven, so that they may be tailored to specific wing geometries and flow conditions. 
In particular, time series data of lift and wing deformation are used to build a model without requiring information about the material properties or mass distribution of the wing. 
Extending the model to predict wing deformation is key for control in aeroelastic applications, to reduce material fatigue due to vibrations or structural failures due to large stresses. 
The ERA algorithm is used to augment the limited measurement data with time delayed information, enabling a reduced order model for fast computation in real-time control. 
We demonstrate this approach on data from a high-fidelity direct numerical simulation of a flexible wing in a low Reynolds number flow.  
Specifically, this model is used for model predictive control (MPC) of an aggressive lift trajectory, while simultaneously minimizing deflection. 
Animals achieve impressive maneuverability and stability with limited computation \cite{Taylor2007}, and one goal of low-order models is to enable similar performance in engineered vehicles.
This modeling framework fills a need for control-oriented models for flying or swimming animals and for MAVs, both of which rely on viscous and unsteady forces.

\section{Background}
Here we provide a brief overview of some of the most relevant background material on data-driven unsteady aerodynamic modeling and control. 
Section \ref{sec:TheoBackground} introduces Theodorsen's model and describes a data-driven extension for a rigid plate~\cite{Brunton2014}, which serves as the foundation for the aeroelastic modeling procedure developed in this paper. This modeling procedure uses the eigensystem realization algorithm, which is described in section~\ref{sec:ERA}. Finally, model predictive control is introduced in section~\ref{sec:MPC} because it is well suited to aeroelastic control due to its ability to constrain deformation. 

\subsection{Empirical Theodorsen model: a data-driven approach} \label{sec:TheoBackground}
Two of the earliest and most important unsteady aerodynamic models for the lift on a flat plate in response to motion are those developed by Wagner~\cite{Wagner1925} in 1925 and Theodorsen~\cite{Theodorsen1935} in 1935. 
These models, which would later be shown to be equivalent~\cite{Garrick1938,Jones1938}, describe the unsteady lift on a two-dimensional pitching and/or plunging flat plate in an incompressible fluid. 
Both models analytically represent the unsteady lift as a function of angle of attack, added mass effects, and the vorticity of an idealized planar wake.

Theodorsen's model for thin airfoils describes the forces and moments on an airfoil for harmonic motions. Theodorsen's model assumes attached flow over the wing with an idealized planar wake in an irrotational, incompressible flow, with no structural damping in the wing. 
For pure pitching motion with dimensional frequency $f$, the contributions to the lift coefficient $C_L$ from both circulation and added mass can be determined as a function of the reduced frequency, $k = \pi f c/U_{\infty}$, the pitch axis, $a$, and the angle of attack $\alpha$ and its derivatives $\dot{\alpha}$ and $\ddot{\alpha}$: 
\begin{equation}
C_L = \frac{\pi}{2} \bigg(\dot{\alpha} - \frac{a}{2} \ddot{\alpha}\bigg) + 2\pi \bigg(\alpha + \frac{\dot{\alpha}}{2}\bigg(\frac{1}{2} - a\bigg)\bigg) C(k).
\end{equation}
Lengths are nondimensionalized by the chord length, $c$, velocities are nondimensionalized by the free stream velocity, $U_{\infty}$, and $a$ represents the pitch axis as a number from $-1$ (leading edge) to $1$ (trailing edge). Theodorsen's transfer function $C(k)$ is defined as
\begin{subequations}
\begin{align}
    C(k) &= \frac{H_1^{(2)}(k)}{H_1^{(2)}(k) + i H_0^{(2)}(k)};\\
    H_v^{(2)} &= J_v - i Y_v,
\end{align}
\end{subequations}
where $J_v$ and $Y_v$ are Bessel functions of the first and second kind, respectively. 
See Leishman~\cite{Leishman2006} for a more complete description of Theodorsen's model.

For control applications, state-space models in terms of ordinary differential equations are preferable \cite{Dowell2016Book,Skogestad2005Book}.
There are numerous examples in the literature of state-space representations of Theodorsen's and Wagner's models~\cite{Jones1938,Vepa1977,Brunton2013_JFS}. 
The coefficients in Theodorsen's model, in particular the $2\pi$ coefficient for the quasi-steady lift, become inaccurate in regimes with significant viscous effects~\cite{Brunton2013_JFM}. 
More accurate coefficients can be obtained from experimental data, as is done in the state-space modeling method developed for rigid airfoils~\cite{Brunton2013_JFS,Brunton2014}:
\begin{subequations}\label{eqn:rigidModel}
\begin{align}
    \frac{d}{dt}\begin{bmatrix} \xv \\ \alpha \\ \dot{\alpha} \end{bmatrix}
    &= \begin{bmatrix} \Av & \boldsymbol{0} & \Bv \\ 
    \boldsymbol{0} & 0 & 1 \\
    \boldsymbol{0} & 0 & 0 \end{bmatrix}
    \begin{bmatrix} \xv \\ \alpha \\ \dot{\alpha} \end{bmatrix}
    + \begin{bmatrix}  \boldsymbol{0} \\ 0 \\ 1 \end{bmatrix} \ddot{\alpha},
    \label{eqn:ss_rigid}\\
    C_L &= 
    \begin{bmatrix}
    \Cv & C_\alpha & C_{\dot{\alpha}} 
 \end{bmatrix} 
 \begin{bmatrix} \xv \\ \alpha \\ \dot{\alpha} \end{bmatrix} 
 + C_{\ddot{\alpha}}  \ddot{\alpha}.
\label{eqn:Observables_rigid}
\end{align}
\end{subequations}

The model coefficients are empirically determined, with $C_\alpha$ replacing the $2\pi$ quasi-steady coefficient and the coefficients $C_{\dot{\alpha}}$ and $C_{\ddot{\alpha}}$ representing the contributions from added mass. 
The dynamics of the model states, $\xv$, are found using the eigensystem realization algorithm (ERA).  
These dynamics take the place of Theodorsen's transfer function, describing the unsteady transient behavior of the fluid wake. The effect of the pitch axis, $a$, is captured in the pitch velocity coefficient, $C_{\dot{\alpha}}$.

\subsection{Eigensystem realization algorithm} \label{sec:ERA}
ERA generates a linear state-space model from an impulse response, without requiring prior knowledge of the model. 
If enough data is taken so that all transients decay, ERA produces a balanced model for which the observability and controllability Gramians are equal~\cite{Ma2010}. 
To develop an impulse response for a pitching airfoil in direct numerical simulations, a fast smoothed linear step maneuver is implemented in $\alpha$, which may be viewed approximately as a discrete-time impulse in $\dot{\alpha}$.  
The output measurements of interest are denoted by $\Yv$. 
These measurements of the observables of interest, such as lift and drag, are sampled at the same time scale as the length of the impulse, and formed into two Hankel matrices, $\Hv$ and $\Hv'$: 
\begin{subequations}
\begin{equation}
    \Hv = \begin{bmatrix}
    \Yv(t_1) & \Yv(t_2) & \cdots & \Yv(t_{N/2 - 1}) \\
    \Yv(t_2) & \Yv(t_3) & \cdots & \Yv(t_{N/2}) \\
    \vdots & \vdots & \ddots & \vdots \\
    \Yv(t_{N/2 -1}) & \Yv(t_{N/2}) & \cdots & \Yv(t_{N-1})
    \end{bmatrix};
\end{equation}
\begin{equation}
    \Hv' = \begin{bmatrix}
    \Yv(t_2) & \Yv(t_3) & \cdots & \Yv(t_{N/2}) \\
    \Yv(t_3) & \Yv(t_4) & \cdots & \Yv(t_{N/2 + 1}) \\
    \vdots & \vdots & \ddots & \vdots \\
    \Yv(t_{N/2}) & \Yv(t_{N/2+1}) & \cdots & \Yv(t_{N})
    \end{bmatrix},
\end{equation}
\end{subequations}
shown here for a square Hankel matrix.
Each $\Yv(t_i)$ may include measurements of multiple observables, such as lift and drag, at time $t_i$, and they are sometimes referred to as Markov parameters. 
Next, $\Hv$ is decomposed using a truncated singular value decomposition (SVD),
\begin{equation} \label{eqn:SVD}
    \Hv = \Uv \Sigmav \Vv^* = \begin{bmatrix} \tilde{\Uv} & \Uv_T\end{bmatrix} 
    \begin{bmatrix} \tilde{\Sigmav} & \mathbf{0} \\ \mathbf{0} & \Sigmav_T \end{bmatrix} \begin{bmatrix} \tilde{\Vv}^* \\ \Vv^*_T\end{bmatrix},
\end{equation}
where $\Sigmav_T$ are the truncated singular values, and $\tilde{\Sigmav}$ are the largest singular values. The rank of the model can be chosen based on the singular value spectrum $\Sigmav$, or based on the model error on a validation data maneuver.   
The rank-reduced SVD decomposition is used, along with $\Hv'$, to determine the reduced order, discrete-time state-space model of the system,
\begin{subequations}
\begin{align}
    \xv_{k+1} &= \Av \xv_{k} + \Bv \uv_k, \\
    \mathbf{y}_k &= \Cv \xv_k + \Dv \uv_k,
\end{align}
\end{subequations}
where the system matrices are given by
\begin{subequations}
\begin{align}
    \Av &= \tilde{\Sigmav}^{-1/2} \tilde{\Uv}^* \Hv' \tilde{\Vv} \tilde{\Sigmav}^{-1/2} ;\\
    \Bv &= \tilde{\Sigmav}^{1/2} \tilde{\Vv}^* \begin{bmatrix}
    \mathbf{I}_q \\ \mathbf{0}
    \end{bmatrix};\\
    \Cv &= 
    \begin{bmatrix}
    \mathbf{I}_p & \mathbf{0}
    \end{bmatrix}  \tilde{\Uv} \tilde{\Sigmav}^{1/2},
\end{align}
\end{subequations}
and $\Dv$ is the first Markov parameter in $\Hv$.  
Here, $q$ is the number of control inputs and $\mathbf{I}_q$ is the $q \times q$ identity matrix; $p$ is the number of system outputs and $\mathbf{I}_p$ is the $p \times p$ identity matrix.

\subsection{Model predictive control} \label{sec:MPC}
In this work, we will use our unsteady aeroelastic models for model predictive control (MPC) to track aggressive lift maneuvers while minimizing wing bending.  
Model predictive control is a widely-used control optimization technique, due to the ability to handle uncertain and nonlinear dynamics and to include constraints. 
Each control action is determined based on minimizing a highly customizable cost function over a receding predictive horizon. Both input and output constraints can be incorporated into the control optimization. In the context of flight, this means that actuation can be constrained to match motor specifications, angle of attack can be constrained to avoid stall, and wing deformation can be constrained to avoid structural damage.

A drawback of MPC is an increase in on-board computation, which can be mitigated by the use of low-order and linear models. As with all model-based control, accurate models enhance performance. Some key parameters for achieving accurate control while limiting the computational cost are the prediction and control horizons, which determine how many time steps in the future the system output and control inputs are calculated, respectively. The control horizon, $T_c = m_c \Delta t$, is generally less than or equal to the prediction horizon, $T_p = m_p \Delta t$. The cost function used to find the optimal control sequence, $\mathbf{u}(\mathbf{x}_j) = \{\mathbf{u}_{j+1},...,\mathbf{u}_{j+m_c}\}$, for the current state estimate or measurement, $\mathbf{x}_j$, over the prediction horizon is

\begin{equation}
    J = \sum_{k=0}^{m_p-1} \| {\mathbf{x}}_{j+k} - \mathbf{r}_{k} \|_{\mathbf{Q}}^2  + \sum_{k=1}^{m_c-1} (\| \mathbf{u}_{j+k} \|_{\mathbf{R}_u}^2 + \| \Delta \mathbf{u}_{j+k} \|_{\mathbf{R}_{\Delta u}}^2 )\,, 
\label{eqn:costFcn}
\end{equation}
subject to constraints. The weighting matrices $\textbf{Q}$, $\textbf{R}_u$, and $\textbf{R}_{\Delta u}$, are used to prioritize the outputs, inputs, and rate of change of the inputs, respectively.

\section{Low-dimensional aeroelastic modeling framework} \label{method}
The model in equation~\eqref{eqn:rigidModel} is formulated for a rigid wing and does not provide predictions of wing deformation. 
Prediction of deformation is essential for active control of flexible wing shapes. Flutter and wing vibrations have been a recognized problem since the early days of flight, and was a key reason for the development of Theodorsen's model. 
The aerodynamics community now recognizes the benefits of flexible structures, especially for increased efficiency, and advances in sensors, actuators, and algorithms are leading to successes in active flutter and vibration control. 

We extend the rigid model in equation~\eqref{eqn:rigidModel} to include wing deformation, resulting in a low-order, linear aeroelastic model. 
Because the rigid model has been widely studied, we seek to minimally modify this structure so that our model can fit naturally into existing model-based control efforts.  
This modeling method does not require any information about the structure, such as the flexural rigidity or mass distribution, as an input. 
The aeroelastic model developed here is designed for wings that passively bend, without any direct control over camber. Instead, wing vibrations can be attenuated using only pitch control in concert with a controller designed based on the linear model.

\subsection{Aeroelastic model identification from an integrated impulse in $\dot{\alpha}$} \label{sec:method_add}
To formulate the model, time series data of $C_L$ and wing deformation are collected after an impulse response in $\dot{\alpha}$. In this work, wing deformation is described for a two-dimensional plate using the local curvature, $\kappa$, near the leading edge. Curvature at other locations could be chosen, based on any areas of particular structural concern or expectations about bending mode shapes, and higher order structural modes may be learned with higher-dimensional measurements along the chord. In an experimental setting or with a 3D wing, strain would generally be used instead of curvature, generally with at least one strain sensor near the wing root. 

As discussed earlier, because of added mass forces, it is impractical to command an actual impulse in $\dot{\alpha}$ in direct numerical simulations or in experiments.  
Instead, a smoothed linear ramp-up in $\alpha$ is commanded over a short time $\Delta \tau_c$, which approximately corresponds to a discrete-time delta input to $\dot\alpha$. 
An example of the smoothed linear ramp-up maneuver can be seen in the appendix, Fig.~\ref{fig:input_data}, which is modified from the Eldredge maneuver~\cite{eldredge:2009,OL2010}. 
The measurements are integrated to obtain the response to an approximate impulse in $\ddot\alpha$. 
Using $C_L$ and $\kappa$ data from the response to an impulse in pitch velocity, a model of the following form is built using a similar procedure as for the rigid unsteady aerodynamic models~\cite{Brunton2014}:
\begin{subequations}
\begin{align}
        \frac{d}{dt}\begin{bmatrix} \xv \\ \alpha \\ \dot{\alpha} \end{bmatrix}
    &= \begin{bmatrix} \Av & \boldsymbol{0} & \boldsymbol{0} \\ 
    \boldsymbol{0} & 0 & 1 \\
    \boldsymbol{0} & 0 & 0 \end{bmatrix}
    \begin{bmatrix} \xv \\ \alpha \\ \dot{\alpha} \end{bmatrix}
    + \begin{bmatrix}  \Bv \\ 0 \\ 1 \end{bmatrix} \ddot{\alpha},
    \label{eqn:ss_add}\\
        \begin{bmatrix} C_L \\ \kappa \end{bmatrix} &= 
    \begin{bmatrix} \vert & \vert & \vert \\
    \Cv & C_\alpha & C_{\dot{\alpha}} \\
    \vert & \vert & \vert
 \end{bmatrix} 
 \begin{bmatrix} \xv \\ \alpha \\ \dot{\alpha} \end{bmatrix} 
 + \begin{bmatrix} \vert \\ C_{\ddot{\alpha}}  \\ \vert \end{bmatrix} \ddot{\alpha}.
\label{eqn:Observables_Theo}
\end{align}
\end{subequations}
In particular, the coefficients $C_{\alpha}$, $C_{\dot{\alpha}}$, and $C_{\ddot{\alpha}}$ are identified by isolating specific components of the output response, and the remaining transient lift is modeled by the ODE in $\xv$ using ERA.  
The vector of coefficients associated with the quasi-steady lift and deformation, $C_\alpha= \Yv_N/\Delta \alpha$, is found by dividing the steady state outputs by the magnitude of the step change in angle of attack, $\Delta \alpha$. $\Yv_N$ is the last measurement, taken when transients have largely decayed. $C_{\dot{\alpha}}= \Yv_m/\dot{\alpha}_m$ is found from the moment of maximum impulse in $\dot{\alpha}$, when $\ddot{\alpha} = 0$, at time $\tau_m$. $C_{\ddot{\alpha}}$ and the Markov parameters for ERA are found from the response of an impulse in $\ddot{\alpha}$, which is achieved with integration, $\tilde{\Yv} = \int \Yv d\tau_c$, starting at the point of maximum impulse, $\tau_m$. This integration step has benefits for noise filtering and suppression, which is discussed further in the appendix. The added mass from acceleration is $C_{\ddot{\alpha}} = \Delta \tau_c \tilde{\Yv}_0/\Delta \alpha$, where $\Delta \tau_c$ is the time length of the impulse maneuver. The accuracy of the identified coefficients is improved by using sampling with a finer time step than the Markov parameters during the maneuver, to determine the empirical coefficients. Once the Theodorsen-like coefficients have been determined, ERA is used to identify the remaining transients using the coarsely-sampled Markov parameters. Details of the method described in this section can be found in the appendix as well as a comparison with an alternate method which omits the integration step.

\section{Results} \label{sec:results}
We demonstrate the modeling approach above, specifically the formulation in Section~\ref{sec:method_add}, using data from a high fidelity fluid-structure interaction numerical simulation~\cite{Goza2017}. The data is generated at $Re=100$, with plate length and stiffness parameters that are similar to those of a insect wing. These parameters were chosen to model a system with significant effects from viscosity and wing deformation~\cite{Cheng2011}.

We show that the the low-rank linear model matches the high fidelity simulation for a rapid test maneuver, as long as the model rank is sufficient to capture the important plate bending modes. For model interpretation and control planning purposes, the predicted lift and deformation can be separated into contributions from quasi-steady and added mass effects, as well as the transient contributions from the viscous wake. 

An important use of this aeroelastic model is to enable simultaneous control of unsteady aerodynamic forces and plate bending by actuation with only pitching motions. This is demonstrated in the high-fidelity FSI simulation by tracking an aggressive reference lift, followed by fast attenuation of plate vibrations, with control planning done with the low-rank linear model and model predictive control (MPC).

\subsection{Direct numerical simulation}

We generate training data to develop the low-rank model by performing direct numerical simulation of a flow over a two-dimensional thin deforming plate with a strongly-coupled immersed boundary projection method~\cite{Goza2017}. The fluid-structure interaction system is governed by three dimensionless parameters: Reynolds number $Re = c U_{\infty}/\nu$, mass ratio $M_\rho = \frac{\rho_s h}{\rho_f c}$, and bending stiffness $K_B = \frac{EI}{\rho_f U_\infty^2 c^3}$. Here, $\rho_s$ and $\rho_f$ are the density of the plate and fluid, respectively, $c$ is the chord length of the two-dimensional plate, $h$ is the thickness of the plate, and $\nu$ is the kinematic viscosity. We fix $Re = 100$ and $M_\rho = 3$ for all simulations in this work. Three different bending stiffnesses of varying orders of magnitude are considered with $K_B = \{0.3125, 3.125, 31.25\}$. The bending stiffness values were chosen to be similar to those of the leading and trailing edges of an insect wing~\cite{Combes2003}, to model a system with significant wing deformation. Further details of the numerical simulation can be found in the appendix.

\begin{figure*}
\vspace{-.1in}
 \centering
\includegraphics[width=1\textwidth]{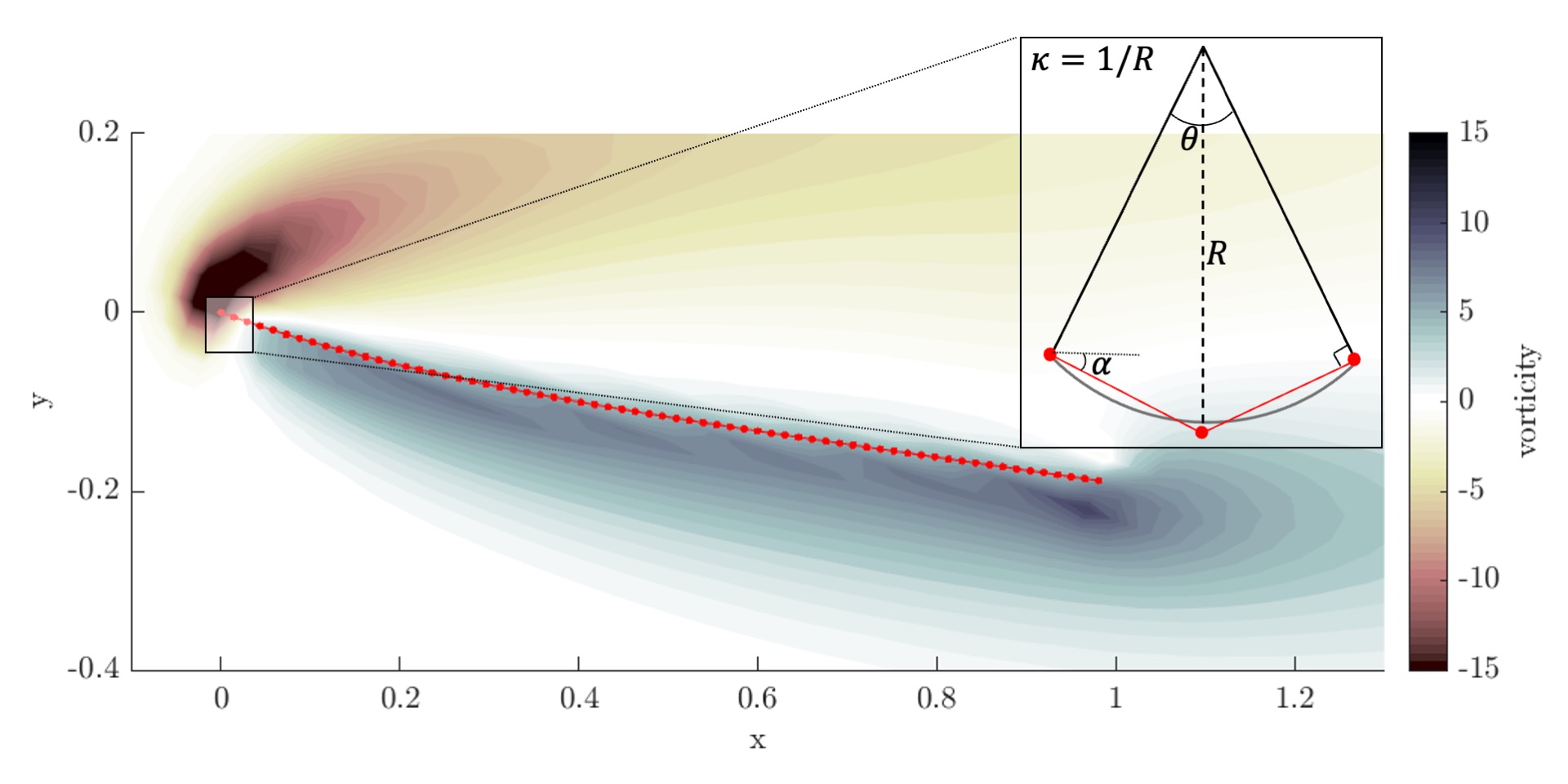}
\caption{The angle of attack, $\alpha$ is measured from the leading edge; in the case illustrated $\alpha = -20^\circ$. Curvature, $\kappa$, is measured from the angle formed by two discrete segments, in this case at the leading edge. There are 66 segments in the plate; red dots indicate the joints between segments. Plate curvature in the inset is exaggerated for illustration.}
\label{fig:plate}
\end{figure*}

Plate deformation is described using curvature, $\kappa$,  $0.015c$ downstream of the leading edge, across an area $0.03c$ in length, shown in Fig.~\ref{fig:plate}. This location was chosen in this study because the two-dimensional plate is pinned at the leading edge, at the pitch point, which results in maximum curvature near this region. The curvature is defined as 
\begin{equation}
    \kappa = \frac{1}{R} = \frac{\theta}{2\Delta c},
\end{equation}
where $\Delta c = 0.015c$ is the length of one discrete segment of the plate. 

\begin{figure*}
\vspace{-.1in}
 \centering
\includegraphics[width=1\textwidth]{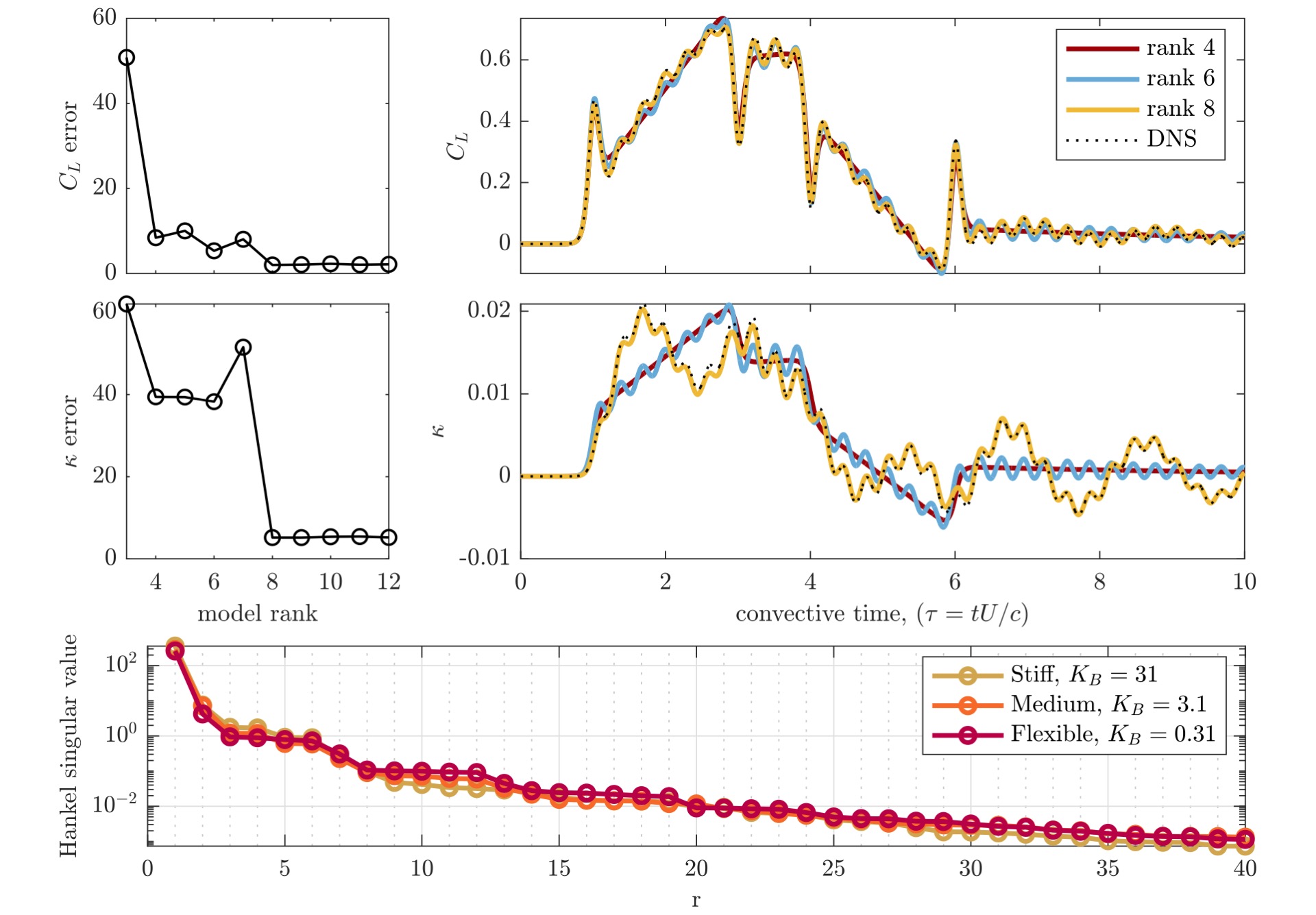}
\caption{Model rank is chosen using a test maneuver, or using the Hankel singular values. (top and middle) Using an aggressive test maneuver, the error can be compared quantitatively (left) or qualitatively (right). When the model order is too low, the quasi-steady and added mass effects are captured, but not the transients associated with the viscous wake and plate bending modes. The models have one state for $\alpha$, one for $\dot{\alpha}$, and the remaining states are associated with transients. The maneuver used for this test can be seen in the bottom panels of Fig.~\ref{fig:contributions}. The data shown in the upper plots is for bending stiffness $K_B = 3.1$. (bottom) If ground-truth data is difficult to obtain, the Hankel singular values from ERA can also be used to choose the model rank. }
\label{fig:rank}
\end{figure*}

\subsection{Model accuracy and comparison with Theodorsen's model} \label{sec:Theo_comparison}

For the systems described in this paper, the low-order linear model shows excellent agreement with the high fidelity simulation with a model rank of roughly eight. The model rank can be chosen from the Hankel singular values, which are the diagonal entries of $\Sigma$ from Eq.~\ref{eqn:SVD}. While there exist theoretically optimal thresholds for rank truncation using singular values \cite{Gavish2014}, comparing with a test maneuver can provide a more interpretable picture of which physics are lost, such as high or low frequency flutter or wake vorticity. In this work, the model rank was chosen by comparison to high fidelity simulation data from an aggressive pitch-up, pitch-down test maneuver~\cite{eldredge:2009,OL2010,Brunton2014}. Both the Hankel singular values and test maneuver rank comparison are shown in Fig.~\ref{fig:rank}.
 This maneuver, and the inclusion of several convective times with no wing actuation, are chosen to balance the effects of errors from fast maneuvers and from transients. The error, $e$, of the output of the reduced order model, $\Yv_{ROM}$, compared to the output of the high fidelity numerical simulation, $\Yv_{DNS}$,  is 
 \begin{equation}
     e = 100 \frac{\sqrt{\sum (\Yv_{DNS} - \Yv_{ROM})^2}}{\sum \Yv_{DNS}^2}.
 \end{equation}
 
\begin{figure*}
\vspace{-.1in}
\centering
\includegraphics[width=1\textwidth]{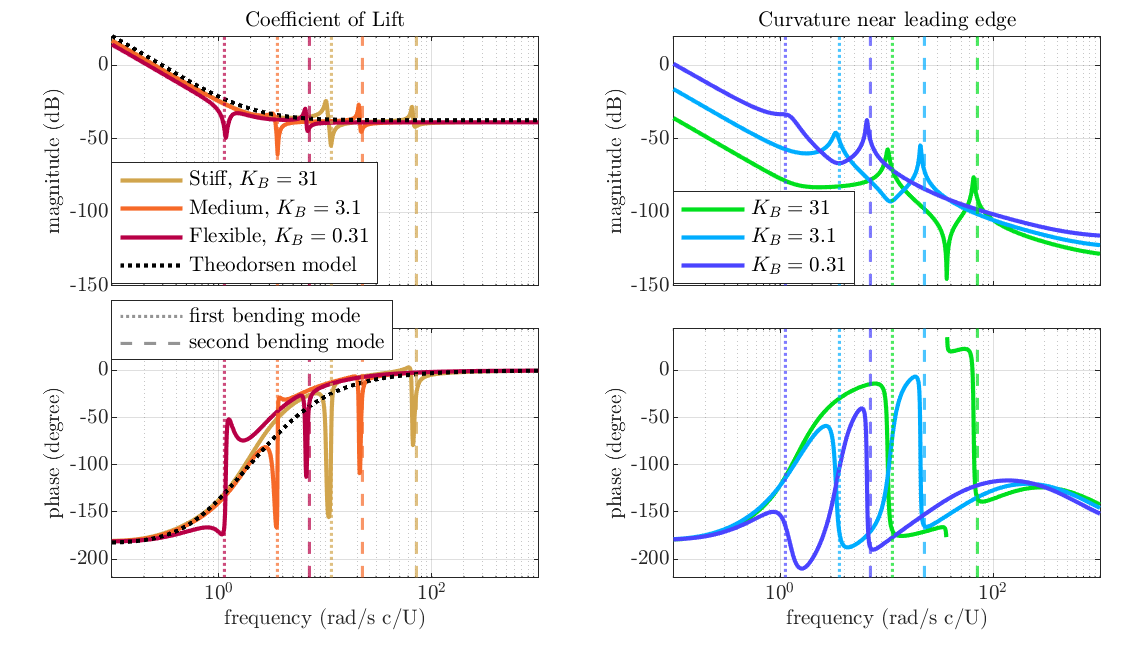}
\caption{Frequency response of lift and deformation from forcing with $\ddot{\alpha}$. The Theodorsen model does not capture resonance behavior from bending modes or vortex shedding. The vertical lines indicating natural frequencies of the bending modes are based on analytically calculated, undamped, bending modes based on the plate dimensions and flexural rigidity.}
\label{fig:Bode}
\end{figure*}

The rank is chosen to capture the most energetic bending modes, which is achieved in the example shown in Fig.~\ref{fig:rank} with a rank of eight: six states to represent fluid transients and plate bending, and one state each for $\alpha$ and $\dot{\alpha}$. A frequency response plot, shown in Fig.~\ref{fig:Bode}, indicates that two bending modes are captured for this system. The models shown in the frequency response plot were developed using the method in Section~\ref{sec:method_add}. Models developed using the alternate method, described in the appendix, can capture additional bending modes if they are present in the signal, but are prone to overfitting if the chosen rank is too high, due to the higher-variance frequency content of the signal prior to integrating. 

The empirically determined model captures resonance responses that are not captured with Theodorsen's model. For context, the first two natural frequencies of an undamped beam matching the plate properties were analytically calculated and are shown in the vertical lines in Fig.~\ref{fig:Bode}. The analytically calculated natural frequencies do not exactly match the model generated from data, likely due to fluid damping, although they are quite close. An advantage of the empirical models shown is they do not require knowledge of the fluid damping or the wing's natural frequencies to generate an accurate model.

\subsection{Model interpretation} \label{interpret}
This modeling method allows the contributions from i) quasi-steady effects from $\alpha$, ii) added mass effects due to $\dot{\alpha}$ and $\ddot{\alpha}$, and iii) plate vibrations, wake vorticity, and transients to be analyzed independently of each other. This information can be used to understand the physics driving the system behavior, and to design controllers that take advantage of these physical phenomena. 

\begin{figure*}
\vspace{-.1in}
 \centering
\includegraphics[width=1\textwidth]{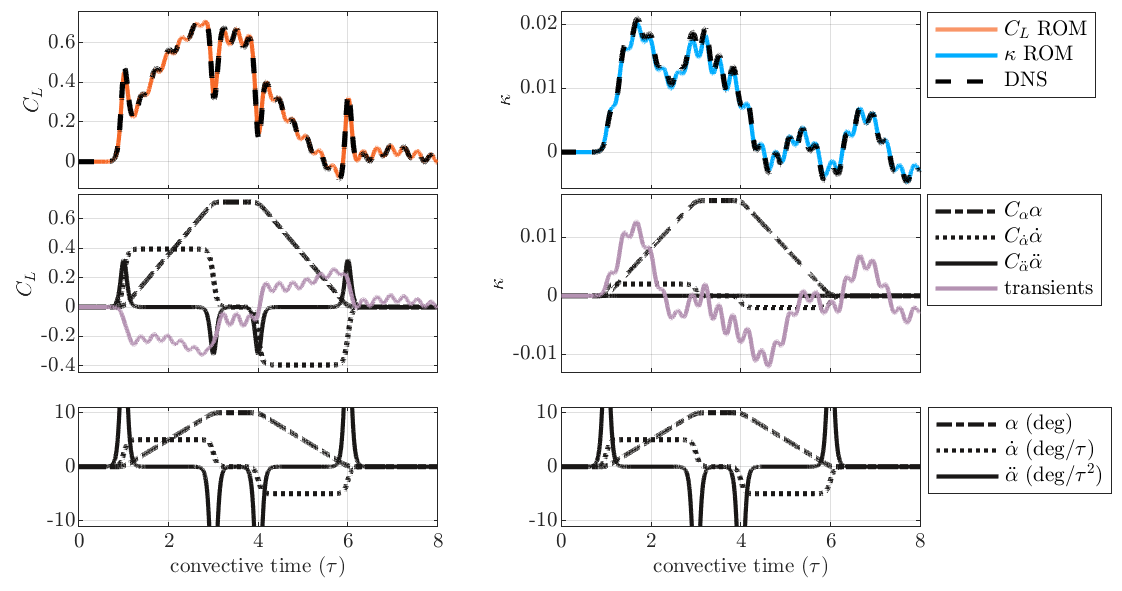}
\caption{Contributions to lift and curvature from pitch angle, velocity, acceleration, and transients. In the top panels, the complete model is shown, and agrees well with the DNS. The middle panels show the contributions from each component of the model. The bottom panels show the test maneuver used. The data shown is for $K_B = 3.1$, with reduced order model rank of 9 (7 transient states). }
\label{fig:contributions}
\end{figure*}

Fig.~\ref{fig:contributions} shows these contributions for the test case described in Section~\ref{sec:Theo_comparison}. The most important contributions to leading edge curvature are $\alpha$ and the latent states representing transients and bending modes, while $\dot{\alpha}$ and $\ddot{\alpha}$ contribute relatively little. This can also be intuited by looking directly at the sign and magnitudes of the empirically determined $\Cv$ matrix, shown below for the model used in Figs.~\ref{fig:rank} and~\ref{fig:contributions}, with $\alpha$ in degrees:
\begin{equation}
    \begin{bmatrix} {C_L} \\ {\kappa} \end{bmatrix} =
    \begin{bmatrix} 
    \Cv_{C_L} & \num{7.1e-2} & \num{7.9e-2} \\
     \Cv_{\kappa} & \num{1.6e-3}  &  \num{4.0e-4}
 \end{bmatrix} 
 \begin{bmatrix} \xv \\ \alpha \\ \dot{\alpha} \end{bmatrix} 
 + \begin{bmatrix}   \num{1.2e-2}  \\  \num{-6.1e-7} \end{bmatrix} \ddot{\alpha}.
 \label{eqn:contributions}
\end{equation} 
For $C_L$, all of the components have important contributions. In this example, the contributions from $C_{\dot{\alpha}}$ may be inflated, but the modeling procedure is robust and adjusts for this in the full model by decreasing the contribution due to transients and wing bending. The overestimate of $C_{\dot{\alpha}}$ may be due to lift enhancement from transient plate bending at the point of maximum $\dot{\alpha}$ in the data used to develop the model. 

\subsection{Model demonstration with feedback control} \label{sec:control}

\begin{figure*}
\vspace{-.1in}
 \centering
\includegraphics[width=1\textwidth]{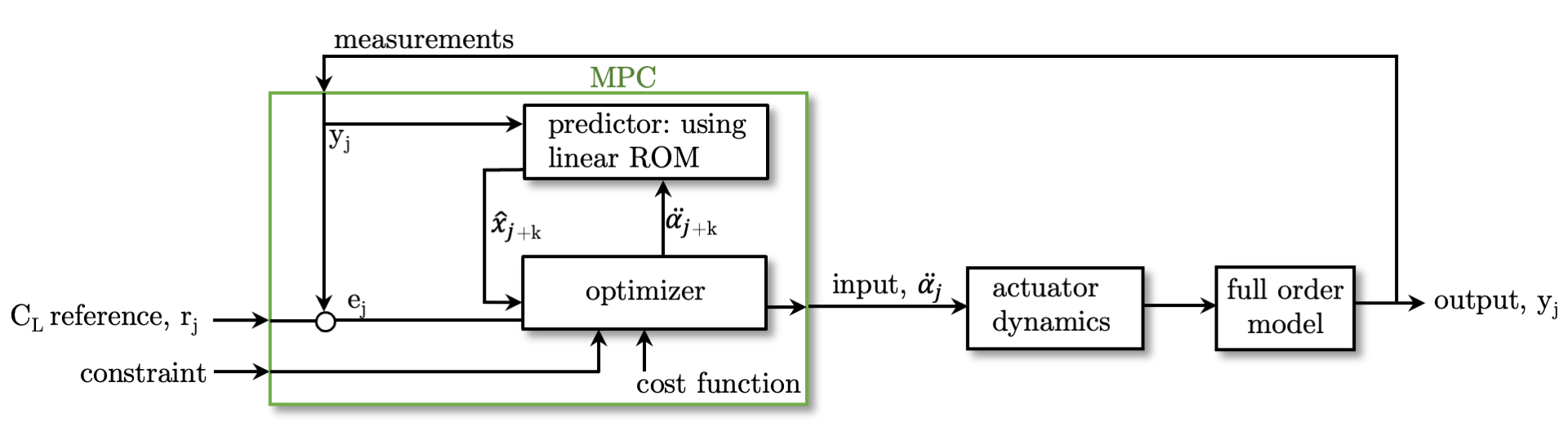}
\caption{Schematic of model predictive control implementation. The predictions inside the control optimization loop are done using the linear reduced order model. The control actions are then applied to a full order fluid-structure interaction simulation.}
\label{fig:control_schematic}
\end{figure*}

To demonstrate how the reduced order aeroelastic models can aid in controller design, feedback control was used to track an aggressive reference lift trajectory while minimizing  wing deformation. Model predictive control is chosen because it is straightforward to implement constraints, for example on $\kappa$ or $\alpha$, and to prioritize either reference tracking of lift or attenuation of wing structural oscillations. In this section, we demonstrate that modeling $\kappa$ is essential for damping structural oscillations, and discuss how the model's empirical coefficients and structure can be used to interpret errors in reference tracking. 

A linear state-space model was first built from step-response data from a numerical simulation, using the procedure described in Section \ref{sec:method_add}, resulting in the most flexible model shown in Fig.~\ref{fig:Bode}, for $K_B = 0.31$ and $r=9$. The control actions were planned based on linear, reduced order state-space models, and were applied to the full-order fluid structure interaction simulation, shown schematically in Fig.~\ref{fig:control_schematic}. The MPC code used is based on that developed by Kaiser et al.~\cite{Kaiser2018}.  
The prediction and control horizons were both 20 time steps, $T_p = T_c = 20 \Delta \tau$, with $\Delta \tau = 0.02$. A Kalman filter was used to estimate the states. Because real motors cannot respond instantaneously, actuator dynamics $G_a = 500/(s + 500)$ were included in the feedback loop, where $s$ is a Laplace domain variable. 

\begin{figure*}
\vspace{-.1in}
\centering
\includegraphics[width=1\textwidth]{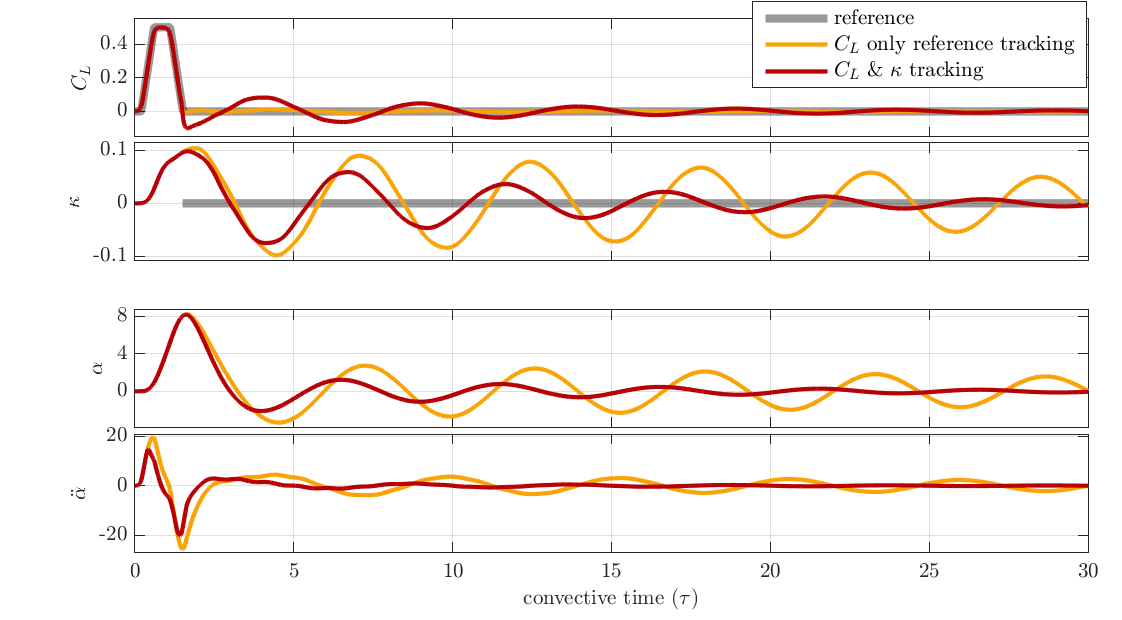}
\caption{Lift and deformation reference tracking.
The MPC actions are planned based on linear, reduced order state-space models, and are applied to a full-order direct numerical simulation. Two control cases are shown: i) tracking a reference value for only $C_L$, and ii) tracking reference $C_L$ and $\kappa$. The reference $\kappa$ shown is only applied to the second case. Angle of attack in degrees, $\alpha$, and actuation, $\ddot{\alpha}$, are shown for reference.}
\label{fig:control}
\end{figure*}

A reference value for the wing deformation can only be specified and tracked if the wing deformation is predicted by the model. Fig.~\ref{fig:control} shows two cases: i) tracking only $C_L$, with no penalty for $\kappa$ oscillations, and ii) tracking a reference $C_L$ and $\kappa$, with the $\kappa$ tracking starting after an impulse in $C_L$. The cost function in equation~\eqref{eqn:costFcn} was used with $\Rv =  0.001$ and
\begin{align}
    \Qv_1 = \begin{bmatrix}
    1 & 0 \\ 0 & 0
    \end{bmatrix}\quad \text{and}\quad
    \Qv_2 = \begin{bmatrix}
    1 & 0 \\ 0 & 10
    \end{bmatrix},
\end{align}
where $\Qv_1$ is used during periods with non-zero reference $C_L$, and $\Qv_2$ is applied for the second case to minimize structural vibrations when the magnitude of the reference $C_L$ is zero. The first case, where only a reference for $C_L$ is specified, shows that wing vibrations from rapid maneuvers persist for a long time after the maneuver, despite the well-controlled lift. 
When a reference deformation is specified, vibrations are quickly attenuated. There is some trade off in lift tracking performance when deformation is included in the MPC cost function. This model makes it possible to tune the optimization weights and control constraints to decide the right trade-off for a given control application, rather than relying only on lift tracking.

\begin{figure*}
\vspace{-.1in}
\centering
\includegraphics[width=1\textwidth]{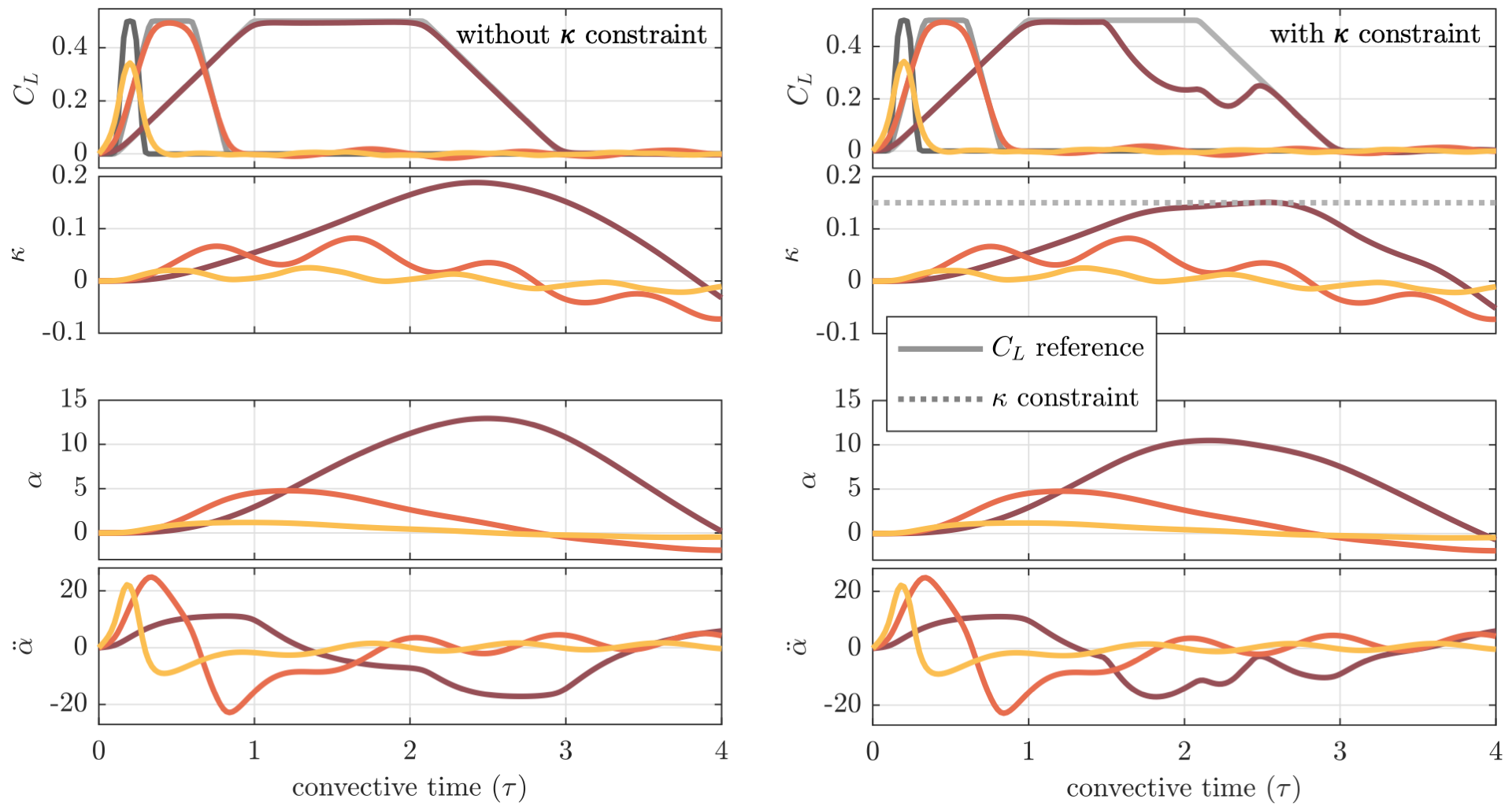}
\caption{The maximum wing deformation depends strongly on the duration of the reference $C_L$ step. (left) Reference $C_L$ is increased, held, and decreased with three different rates of increase and hold time. If the hold time is too short, the system is unable to respond quickly enough. As the hold time increases, the angle of attack, $\alpha$, must also increase, which results in higher maximum curvature, $\kappa$. 
(right) By choosing reference $C_L$ with short enough hold times, moderate constraints on maximum $\kappa$ do not affect reference tracking accuracy. }
\label{fig:constraint1}
\end{figure*}

Limiting the maximum wing deformation is another goal of deformation control, to avoid damage from large stresses. Similar to the model described in Fig.~\ref{fig:contributions} and equation~\eqref{eqn:contributions}, the model used in this section has significant contributions to $C_L$ from $\ddot{\alpha}$, due to added mass; however, the largest contribution to $\kappa$ is from $\alpha$. Shown in Fig.~\ref{fig:constraint1} (left), this results in small deformations when the reference $C_L$ is stepped-up and then stepped back down in a short time, because the MPC optimization is able to use added mass forces to generate most of the increase in $C_L$. 
However, if the high reference $C_L$ is  held for longer, there is no longer added mass due to acceleration, and the angle of attack must increase to sustain the $C_L$ at the reference value. 
When the reference maneuver is too rapid, the controller is unable to respond quickly enough.  

\begin{figure*}
\vspace{-.1in}
 \centering
\includegraphics[width=1\textwidth]{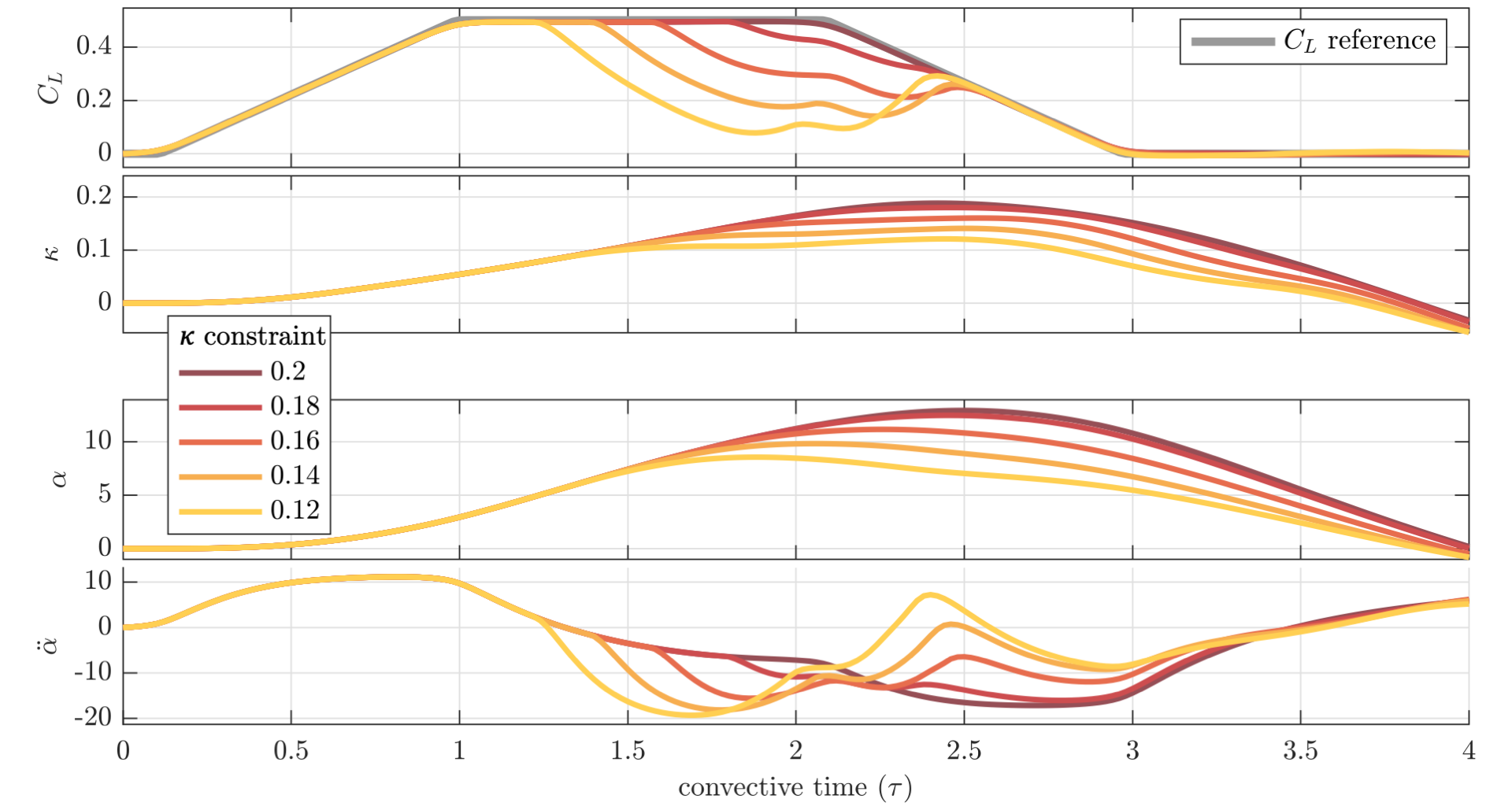}
\caption{For systems with significant added mass effects on $C_L$, constraints on deformation decrease performance when $C_L$ is held constant, but does not affect accuracy at the beginning of maneuvers. }
\label{fig:constraint2}
\end{figure*}

An understanding of the relative contributions from added mass and angle of attack can be used to balance the need for constraining deformation with desired trajectories. In Fig.~\ref{fig:constraint1} (right), the $\kappa$ constraint does not affect shorter maneuvers, but leads to significant error for a longer maneuver. In this regime, constraining $\kappa$ effectively also constrains $\alpha$. As constraints on deformation become more aggressive, the lift of the controlled system falter earlier, as shown in Fig.~\ref{fig:constraint2}.

\section{Discussion}

In this work, we describe a method for obtaining accurate low order, linear state-space aeroelastic models for control. The method uses lift and deformation data from an impulse response to construct the model, providing accurate predictions without requiring information about the surrounding flow field or wing structural properties. Interpretable coefficients relating to added mass, lift slope, and transient effects are built in to the model, providing insights about the underlying physics. The remaining dynamics are modeled using ERA, which accurately captures the transients dynamics due to the viscous wake. The resulting model is low dimensional and captures the dominant dynamics, extending rigid state-space aerodynamic models~\cite{Brunton2014} to account for wing flexibility. 

These models are well-suited for use with standard control techniques, which is demonstrated using MPC to track an aggressive reference lift while attenuating oscillations in leading edge curvature for an insect-inspired two-dimensional wing at $Re = 100$. These state-space models allow analysis and control design in both the time and frequency domains, including identification of resonant frequencies of the flexible structure. Because the wing deformation can be predicted, wing vibrations or flutter can be actively controlled by modulating only the angle of attack. 

Curvature was used as the measure of deformation in this paper because the data was based on a simulation without a well-defined plate thickness. In real-world and experimental settings, strain may be a more appropriate deformation observable due to ease of measurement. Expanding the modeling algorithm to include a preprocessing step of determining the optimal strain sensor location may be of value~\cite{Mohren2018pnas}. Another possible extension of this modeling method is to include coefficient of thrust as an observable, giving a more complete picture of the forces on the wing, making the work more relevant for flapping and hovering flight. 

The demonstration of this modeling framework at $Re = 100$ and mass ratio of $M_\rho = 3$ was particularly motivated by development of MAVs and to investigate control strategies used in animal flight, due to the inclusion of viscous effects. Managing wing loading while increasing efficiency at small scales is an ongoing challenge, which these models are designed to address. Further work is necessary to demonstrate the method presented for higher $Re$, applicable to wind turbines or aircraft, and for realistic 3D wings. For these applications, inclusion of multiple deformation or strain sensors is likely to be advantageous to capture relevant wing bending modes. For aeroelastic structures with low stiffness and high mass ratio, the wake may be irregular and transient wing deformations may no longer be periodic, limiting the effectiveness of this modeling procedure. It will also be important to extend these models to handle larger amplitude maneuvers, either by combining multiple linear models generated from maneuvers at several angles of attack using gain scheduling or LPV models~\cite{Hemati2017}, or by including nonlinear terms in the model.

\section{Appendix}
\subsection{Detailed procedure for obtaining model} \label{sec:model_method}
\begin{figure*}
\vspace{-.1in}
 \centering
\includegraphics[width=1\textwidth]{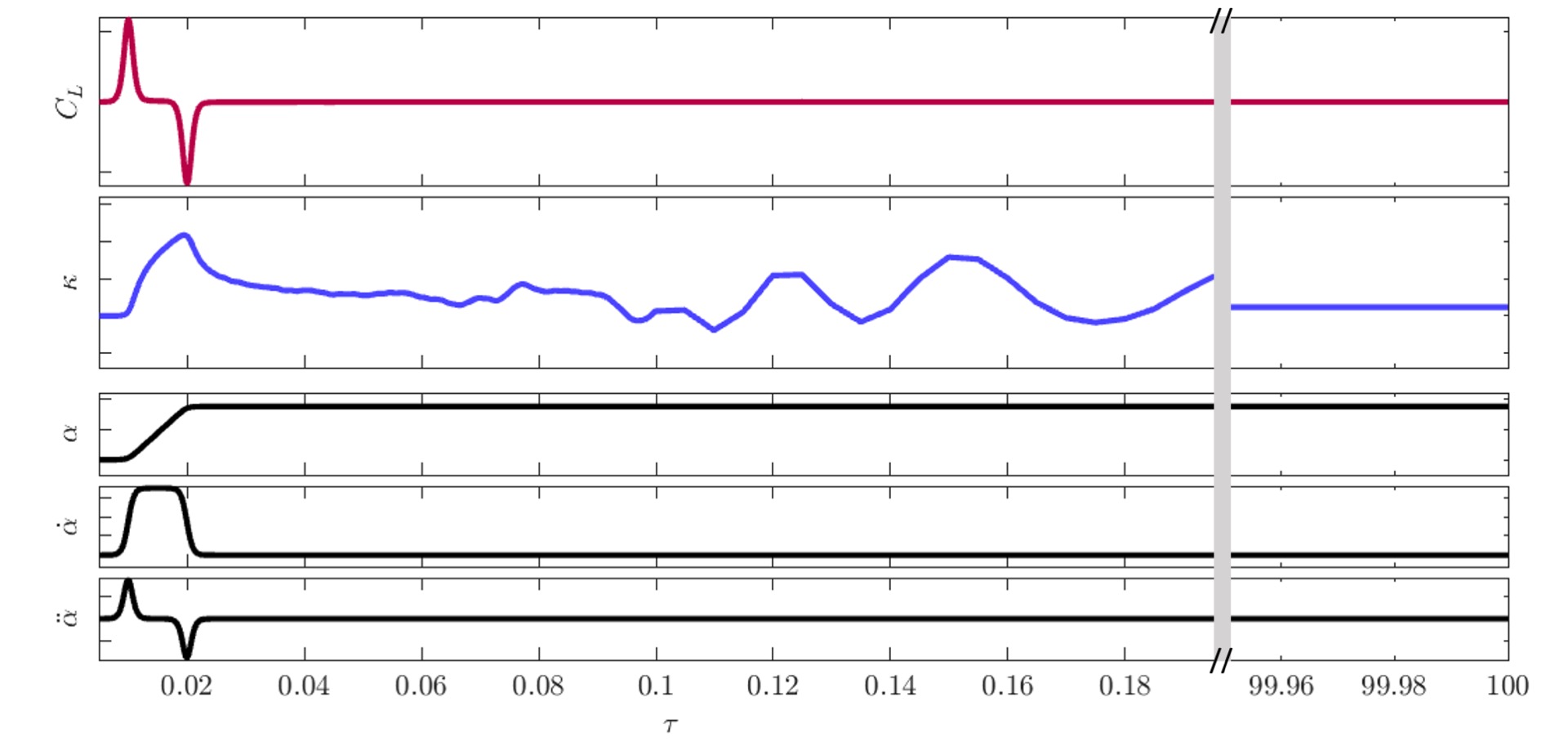}
\caption{An impulse in $\dot{\alpha}$ is used to generate training data used for the system identification algorithm. The data shown is for $K_B = 0.31$, for a short time window during and after the impulse, as well as the end of the time series, when transients and wing bending have ceased.}
\label{fig:input_data}
\end{figure*}

Here we provide a detailed step-by-step procedure to identify unsteady aeroelastic models from data, roughly following the procedure for rigid unsteady aerodynamic modeling~\cite{Brunton2013_JFM,Brunton2014}.  
Training data is in the form
\begin{equation}
    \Yv = \begin{bmatrix}
    \vert & \vert \\
    C_L   & \kappa   \\
    \vert & \vert
    \end{bmatrix}.
\label{}
\end{equation}
An example of the impulsive maneuver and training data is show in Fig.~\ref{fig:input_data}. The initial impulse data is shown, which will generally be only a small portion of the overall time series. Measurements should continue to be taken until the system reaches steady state. To obtain an approximately linear response, a small step amplitude, $\Delta \alpha \in [0.1^\circ , \; 1^\circ]$, and short time, $\Delta \tau \in [0.01, \; 0.1]$, should be used. For experimental data, it may be necessary to employ the observer-Kalman filter identification (OKID) approach to obtain the impulse response~\cite{Juang1991}, which has been demonstrated for the identification of rigid unsteady aerodynamic models from data~\cite{Brunton2013_JFM,Brunton2014}.

The steps to obtain the coefficients in equation~\ref{eqn:Observables_Theo} are: 
\begin{enumerate}
    \item Form the data matrix $\Yv$ from a step response in $\alpha$. Update $\Yv$ by subtracting the initial values, $\Yv_0 = \begin{bmatrix}
    C_L(\tau = 0) & \kappa(\tau = 0)  \end{bmatrix}$ of each observable from the rest of the time series. 
    \item $C_\alpha= \Yv_N/\Delta \alpha$, is found by dividing the steady state by the magnitude of the step change in angle of attack, $\Delta \alpha$. $\Yv_N$ is the last measurement. 
    
    Update $\Yv$ by subtracting the steady-state contribution, $C_\alpha \alpha_k$ from each $\Yv_k$. \label{step2}
    \item $C_{\dot{\alpha}}= \Yv_m/\dot{\alpha}_m$, found from the moment of maximum impulse in $\dot{\alpha}$, when $\ddot{\alpha} = 0$, at time $\tau_m$.
    Update $\Yv$ by subtracting the pitch velocity contribution, $C_{\dot{\alpha}} \dot{\alpha}_k$ from each $\Yv_k$.
    \item $C_{\ddot{\alpha}}$ and the Markov parameters for ERA are found from the response of an impulse in $\ddot{\alpha}$, which is achieved with integration, $\tilde{\Yv} = \int \Yv d\tau_c$, or in practice for a discrete signal, with a cumulative sum, starting at the point of maximum impulse, $\tau_m$. 
    \item $C_{\ddot{\alpha}} = \Delta \tau_c \tilde{\Yv}_0/\Delta \alpha$, where $\Delta \tau_c$ is the time length of the impulse maneuver. 
    \item The discrete-time state-space matrices, $\Av_d$, $\Bv_d$, $\Cv$, $\Dv$, and the states, $\xv$, are found using ERA, with the reduced order rank as an additional parameter. The Markov parameters used for ERA are the coarse-time integrated signal, $\tilde{\Yv}$.
    \item The discrete time matrices must be converted to continuous time before assembling the model in the form in equations~\ref{eqn:Observables_Theo} \& \ref{eqn:ss_add}. If there were non-zero initial conditions, they should be added to the model as a constant term.
\end{enumerate}

All models shown here were generated from an impulse magnitude of $\Delta \alpha = 0.1^\circ$. The size of the Hankel matrix, $\Hv$, was roughly 5,000 x 5,000. The size of the Hankel matrix will depend on the duration of the impulse, and overall duration required for transients to die out.

\subsection{Aeroelastic model identification from an impulse in $\dot{\alpha}$ } \label{sec:method_ad}

\begin{figure*}
\vspace{-.1in}
 \centering
\includegraphics[width=1\textwidth]{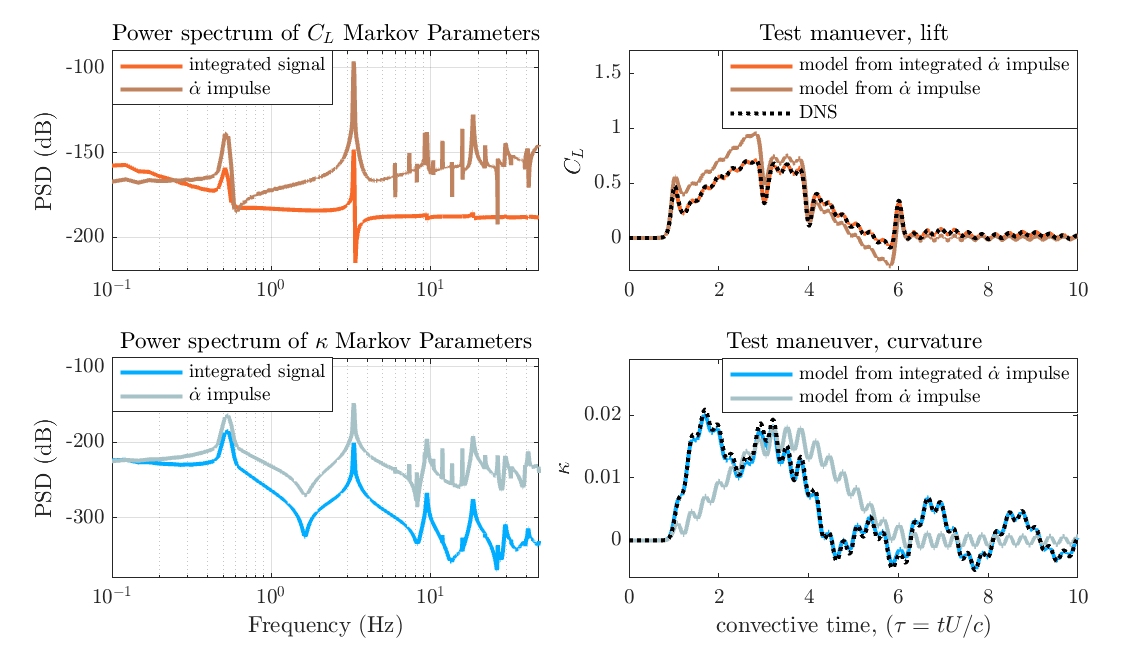}
\caption{The procedure described in Section~\ref{sec:method_add} (integrated signal) generates a more accurate model than the procedure  using an $\dot{\alpha}$ impulse for some cases. For the case shown, $K_B = 3.1$ with rank 9 models, the lowest bending mode is not captured by ERA, resulting in an inaccurate model. (left) Power spectral density (PSD) of the Markov Parameters, $\tilde{\Yv}$. Note the low power in $C_L$ for the $\dot{\alpha}$ impulse for the peak at $0.5$ Hz, the lowest frequency bending mode, compared to the power contained in high frequency noise. (right) The test maneuver shown in Fig.~\ref{fig:rank} is shown again here for each of the two modeling procedure, with rank 9 models, compared with data from the high fidelity DNS model.}
\label{fig:PSD}
\end{figure*}

 Rather than integrating to get an impulse in $\ddot{\alpha}$, the model can be obtained directly from the impulse in $\dot{\alpha}$. This is done by omitting the signal integration step, as in~\cite{Brunton2014}, which results in a model of the form
\begin{equation}
    \frac{d}{dt}\begin{bmatrix} \xv \\ \alpha \\ \dot{\alpha} \end{bmatrix}
    = \begin{bmatrix} \Av & \boldsymbol{0} & \Bv \\ 
    \boldsymbol{0} & 0 & 1 \\
    \boldsymbol{0} & 0 & 0 \end{bmatrix}
    \begin{bmatrix} \xv \\ \alpha \\ \dot{\alpha} \end{bmatrix}
    + \begin{bmatrix}  \boldsymbol{0} \\ 0 \\ 1 \end{bmatrix} \ddot{\alpha},
    \label{eqn:ss_ad}\\
\end{equation}
with the observables in the same form as equation~\eqref{eqn:Observables_Theo}. This alternate method is problematic if there are several orders of magnitude between the spectral power contained in the first several bending modes, or significant noise at coherent frequencies, including due to sampling frequency. This is shown in Fig.~\ref{fig:PSD}, for $K_B = 3.1$, with the same test maneuver shown in Fig.~\ref{fig:rank}. The spectral power contained in the $C_L$ signal for the lowest frequency mode is two orders of magnitude smaller than the power contained in the second mode, and is similar to the power contained in high frequency noise. For the rank 9 model shown, the modeling procedure described in this section fails to capture the lowest frequency mode, resulting in an inaccurate model. By integrating the signal to obtain the Markov parameters, as described in the modeling procedure in Section~\ref{sec:method_add}, the low power bending mode is captured accurately, and the high frequency noise is smoothed, resulting in an accurate model. Increasing the model rank is not a solution to this problem, because the model rank required to capture the missing low amplitude mode generates spurious resonance peaks. However, an advantage to generating the model without integrating is that the model may capture higher frequency bending modes in a low noise signal, which may be lost when integrating the signal. Both methods were previously shown to produce accurate models for rigid systems, where the power spectrum of the Markov Parameters is dominated by vortex shedding rather than structural bending modes.

 To obtain a model in the form of equation~\ref{eqn:ss_ad}, the procedure above in Section~\ref{sec:model_method} is followed through step~\ref{step2}. The remaining steps to obtain the method are described below. 
\begin{enumerate}
\setcounter{enumi}{2}
    \item Identify the time with maximum acceleration, $\tau_n$. $C_{\ddot{\alpha}} = \ddot{\alpha}^{\dagger}_n \Yv_n$, where $\dagger$ indicates the pseudo-inverse. Subtract the pitch acceleration contribution, $C_{\ddot{\alpha}} \ddot{\alpha}_k$ from each $\Yv_k$.
    \item The Markov parameters,$\Bar{\Yv}$, are the signal $\Yv$ sampled at intervals of the coarse time step, $\Delta \tau_c$, starting at the time of maximum impulse, $\tau_m$.
    \item Because the impulse in this version of the method is from $\dot{\alpha}$, $C_{\dot{\alpha}}$ is found from the first Markov parameter; $C_{\dot{\alpha}} = \Delta \tau_c\Bar{\Yv}_0/\Delta \alpha$.
    \item $\Av_d$, $\Bv_d$, $\Cv$, $\Dv$, and $\xv$, are found using ERA with $\Bar{\Yv}$, and then converted to continuous time and assembled into the form in equations~\ref{eqn:ss_ad} \& \ref{eqn:Observables_Theo}. If there were non-zero initial conditions, they should be added to the model as a constant term.
\end{enumerate}

\subsection{Fluid-structure interaction model} \label{sec:DNS}

 We performed direct numerical simulation of a flow over a two-dimensional thin deforming plate with  a  strongly-coupled  immersed  boundary  projection  method. The incompressible Navier--Stokes equation for the fluid was discretized in the vorticity-streamfunction form~\cite{Taira:JCP07} with accurate surface stresses and forces to enforce the boundary condition at the plate~\cite{goza2016accurate}. At the far-field boundaries, uniform flow with free stream velocity $U_\infty$ was prescribed. The solver uses an explicit Adam-Bashforth method and an implicit Crank-Nicolson scheme for discretization of the advective and viscous terms of the Navier--Stokes equation, respectively. This method was validated by Goza and Colonius for a flapping flag in~\cite{Goza2017}.
 
\begin{figure*}
\vspace{-.1in}
 \centering
\includegraphics[width=1\textwidth]{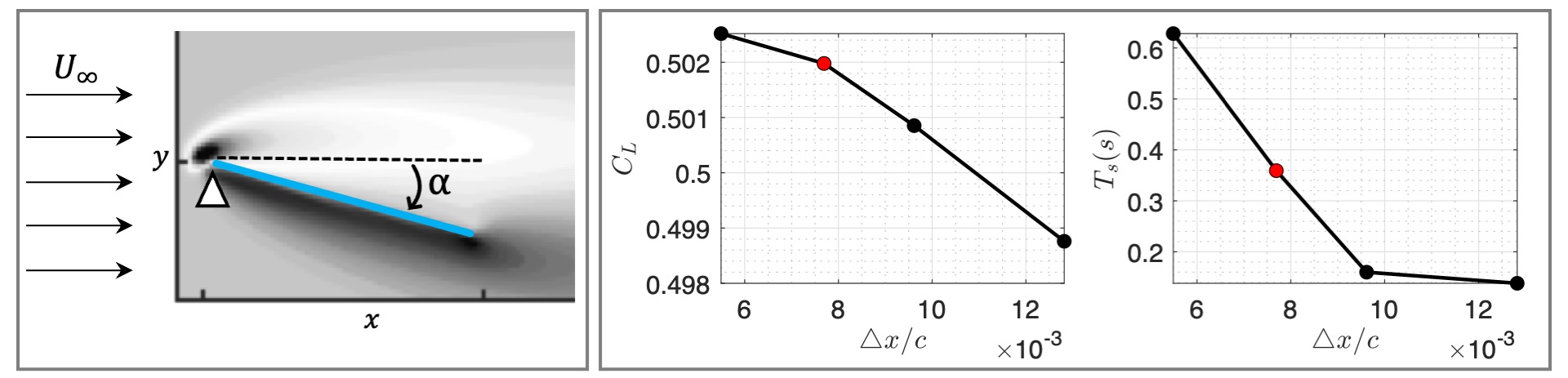}
\caption{Schematic and convergence of direct numerical simulation, shown for an undeformed wing. (left) The angle of attack, $\alpha$, is measured between the incoming free stream flow, and the plate at the leading edge. The pitch axis of the plate is also at the leading edge, where the angle of attack of a deformed or undeformed plate are the same. The full computational domain is not shown in this image. (right) The grid spacing of the finest mesh, shown in red, was chosen as a compromise between accuracy and computational time, $T_s$. A grid spacing of $\Delta x/c \approx 0.0077$, shown in red, error within $0.3\%$ of the steady-state coefficient of lift, $C_L$.}
\label{fig:mesh}
\end{figure*}

 To speed up the computations, a  multi-domain technique with five grid levels was implemented~\cite{Colonius:CMAME08}. The finest domain was fixed at $-0.2 \le x/c \le 1.8, -1 \le y/c \le 1$ with a grid spacing of $\Delta x/c \approx 0.0077$, represented by the red dot in Fig.~\ref{fig:mesh}. Here $c$ is the length of the plate, $x$ is the spatial domain location in the direction of the chord for a plate with no deformation, and $y$ is the spatial domain perpendicular to the chord. The leading edge of the chord, which is the pitch axis, is located at $(0,0)$. This grid spacing was chosen as a compromise between error and computational time, $T_s$, with an error within $0.3\%$ of the steady-state lift coefficient. $T_s$ is the time in seconds to run a single time step. The Reynolds number for all the simulations in this work is $Re \equiv U_\infty c/\nu  = 100$, where $\nu$ is the kinematic viscosity. 

The Euler-Bernoulli equation for the plate was discretized using a co-rotational finite element formulation~\cite{criesfield1991non}.  This formulation enables arbitrary large displacements and rotations by attaching a local coordinate frame to each element. The plate was discretized into $65$ elements with the leading edge placed at $(x/c, y/c) = (0,0)$.

\section*{Data accessibility}
The code used in this work is available at https://github.com/mhickner/aeroelastic-ss-model, including sample data which can be used to generate the models shown in section~\ref{sec:results}.

\section*{Acknowledgements}
The authors acknowledge support from the Air Force Office of Scientific Research (AFOSR FA9550-19-1-0386) and the National Science Foundation AI Institute in Dynamic Systems (Grant No. 2112085). We would like to thank the anonymous referees for their valuable comments that helped strengthen the manuscript.

\printbibliography

\end{document}